\documentclass[12pt]{iopart}
\usepackage{iopams}  
\usepackage{amssymb}
\usepackage{setstack}  
\usepackage{graphicx}
\usepackage{bm}
\newcommand{\Op}[1]{{{\mathrm{\hat{#1}}}}}

\begin{document}

\title[Quantum control with noisy fields]{Quantum control with noisy fields: computational complexity vs. sensitivity to noise}

\author{S. Kallush$^{1}$, M. Khasin$^{2}$  and R. Kosloff$^{3}$ }

\address{$^{1}$Department of Physics, ORT-Braude College, P.O. Box 78, 21982 Karmiel, Israel\\
$^{2}$Nasa Ames Research Center, Moffett Field, CA, USA\\
$^{3}$Fritz Haber Research Center for Molecular Dynamics, 
Hebrew University of Jerusalem, Jerusalem 91904, Israel}
\ead{ronnie@fh.huji.ac.il}

\begin{abstract}
A closed quantum system is defined as completely controllable if an arbitrary unitary transformation can be executed using the available controls.
In practice,  control fields are a source of unavoidable noise, which has to be suppressed to retain controllability.
Can one design control fields  such that the effect of noise is negligible on the time-scale of  the transformation? This question is intimately related to the fundamental problem of a connection between the computational complexity of the control problem and the sensitivity of the controlled system to noise.
The present study considers a paradigm of control, where the Lie-algebraic structure of the control Hamiltonian is fixed, 
while the size of the system increases with the dimension of the Hilbert space representation of the algebra. 
We find two types of control tasks, easy and hard. Easy tasks are characterized by a small variance of the evolving state
with respect to the operators of the control operators. They are relatively immune to noise and the control field is easy to find.
Hard tasks have a large variance, are sensitive to noise and the control field is hard to find. The influence of noise increases
with the size of the system, which is measured by the scaling factor  $N$ of the largest weight of the representation. For fixed time and control field as $O(N)$ for easy tasks and as $O(N^2)$ for hard tasks.
As a consequence, even in the most favorable estimate, for large quantum systems, 
generic noise in the controls  dominates  for a  typical class of target transformations, i.e., 
complete controllability is destroyed by  noise. 
\end{abstract}
\maketitle

\section{Introduction}
\label{sec:intro}

Coherent control aims to steer a quantum system from an initial state to a target state via an external field \cite{rice92,shapiro03,rabitz88,k67}.  The basic idea is to control the interference pathway governing the dynamics.
For given initial and final (target) states the coherent control is termed \textit{state-to-state} control. A generalization is steering simultaneously an arbitrary set of initial pure states to an arbitrary set of final states, i.e. controlling  
a unitary transformation. Such an application sets the foundation for a quantum gate operation \cite{tesch01,k181,rabitz05,glaser2012}.
Controllability of a closed quantum system has been addressed by Tarn and Clark \cite{tarn}. 
Their theorem states that for a finite dimensional closed quantum system,   
the system is \textit{completely controllable}, i.e., an arbitrary unitary transformation of the system can be realized 
by an appropriate application of the controls \cite{rama}, if the  control operators and the unperturbed Hamiltonian  generate the Lie-algebra of all Hermitian operators.  
Complete controllability implies {state-to-state controllability}\cite{dallesandro}.

Experimentally, there has been  remarkable success in constructing devices designed to generate arbitrary control fields \cite{tom99,brixner07,gerber08}.  
Nevertheless, in practice controllability is hard to achieve even for small quantum systems \cite{glaser09,glaser07,herr07}.
Applications toward quantum information processing require  upscaling of the control procedures 
to large quantum systems.  The task of finding a control field for a particular state to state transformation is an inversion problem of high computational complexity \cite{barak07}. 
Some exceptions such as adiabatic methods, allow a constructive approach \cite{mcmahon2008adiabatic}. In this case the field is easy to find
but the implementation may take an infinite amount of time.
The algorithmic task of finding the field that generates
a unitary transformation scales factorially more difficult  with the size of the system\cite{k193}.
This is because a larger number of state-to-state control fields have to be found without interfering with the other fields.  

Even if the control field is found for a closed system, the feasibility of  control depends on the sensitivity of the driven evolution to noise. Real quantum systems are open system subject to noise introduced into the systems dynamics by the system-bath coupling. 
A number of techniques have been designed to combat the  environmental noise and effective and ingenious  
algorithms have been developed and explored \cite{viola99,viola09,lidar05,lidar07,gershon07}. An unavoidable source of noise is the noise originating in the control field. The magnitude of this noise
depends on the properties of the control field. This dependence raises a crucial problem of overcoming the effect of noise with a smarter design of the control field.  
This problem has been extensively investigated in the context of fault-tolerant quantum computation \cite{preskill98,knill2005}, 
where various schemes have been designed to fight the noise in the gates. 

The sensitivity to noise is expected to depend on the computational complexity of the coherent control problem.
In quantum computation the number of gates  increases  with the size of the system. 
In many fields, for example in  NMR  \cite{nielsen10} or in control of molecular systems \cite{rice92} a different control paradigm is standard. There the control operators are fixed while the size of the system may vary.
This is the control paradigm considered in the present study to deal with the two deeply interconnected questions:
\begin{itemize}
\item{Is it always possible, 
for a given target, to design a field, such that the effect of  associated noise can be neglected? }
\item{Is there a connection between computational complexity and the sensitivity to noise of the control solution?}
\end{itemize} 
We consider a quantum system with finite Hilbert space dimension. To be able to scale up,  
we choose the control operators from the elements of the  spectrum-generating algebra of the quantum system \cite{bohm}. 
In this manner, the control Hamiltonian structure is independent of  size. 
The size of the system is determined by the dimension of the Hilbert-space representation of the algebra. 
It can be characterized by a parameter $N$, which is a scaling factor of the highest weight of the representation  \cite{gilmorebook,Gilmore79}. The system is assumed to be completely controllable in the absence of noise.
 In Section II it is shown that complete controllability of a large system, i.e., for $N \gg 1$,
even in the weaker sense of state-to-state control  does not endure in the presence of generic noise in the control fields. A novel scaling relation between an upper bound on the tolerable noise strength and the size of the system is obtained, which shows that the noise strength must decrease at least linearly with the size of the system. As a result, in practice the control is not scalable with the size of the system.  In Section III we present a numerical analysis of various control problems with noise in the single mode Bose-Hubbard model, associated with $su(2)$ spectrum-generating algebra. It is demonstrated that the sensitivity to noise increases with the computational complexity of the control problem. The scaling relations between the size of the system and the loss of fidelity are verified. Section IV includes discussion and conclusions.

\section{The control problem}
\label{sec:prob}

Our target of control is the state-to-state transformation $| \psi_i \rangle \rightarrow | \psi_f \rangle$ in a control time $T$.
The control Hamiltonian has the form:
\begin{equation}
\Op H=\Op H_0+\sum_k \left[u_k(t)+ \xi_k(t)\right]\Op X_k.
\label{eq:1}
\end{equation}
where $u_k(t)$ are control fields, $\Op X_k$ control operators and $\Op H_0$ the stationary drift Hamiltonian. 
The presentation is restricted to a quantum system with finite Hilbert space dimension. It is assumed that the control operators are elements of the  spectrum-generating algebra associated with $\Op H_0$ \cite{bohm}. For a finite dimensional system it is sufficient to consider a compact semisimple algebra \cite{gilmorebook}. The size of the system increases with the dimension of the Hilbert-space representation of the algebra. It can be characterized by a parameter $N$, which is a scaling factor of the highest weight of the representation  \cite{Gilmore79}. The physical interpretation of $N$ depends on the system. It can be, for example, a number of particles in the system or the number of energy levels. From now on we shall denote $N$ as the size of the system.

In the absence of noise the system governed by Eq. (\ref{eq:1}) is chosen to be completely controllable \cite{tarn,rama}.
Generically unavoidable noise is included in the controls $\xi_k(t)$. 
This phenomena is modeled by a delta-correlated Gaussian noise: 
$\left\langle \xi_k(t)\xi_l(t')\right\rangle = 2 \Gamma_k(t) \delta_{kl} \delta(t-t')$. 
Typically, the strength of the noise  $\Gamma_k(t)$  depends on the amplitude of the control field $\Gamma_k(t)= f(u_k(t))$.

The equation of motion for the density operator of the system for this noisy system is given by \cite{gorini76}:
\begin{eqnarray}
\frac{\partial}{\partial t}\Op \rho &=&-i \left[\Op H_0+\sum_k u_k(t)\Op X_k, \Op\rho \right] \nonumber \\
&-& \sum_k  \Gamma_k(t)\left[\Op X_k,\left[\Op X_k,\Op\rho \right]\right]. \label{liouville}
\end{eqnarray}
The Gaussian dissipative term represented by a double commutator is responsible for loss of coherence.
How are the control objectives influenced by  the unavoidable noise?

\subsection{Purity and fidelity}

Due to  noise an initial pure state will degrade. The purity measured by ${\cal P} \equiv \bf{Tr}\left\{ \Op \rho^2\right\}$ of an initially pure state $\Op \rho=\left|\psi\right\rangle\left\langle \psi\right|$   will decrease. For a noisy control the state to state controllability is preserved  only if the purity loss during the target transformation is negligible, i.e., $1- {\cal P} \ll 1$.  
In fact, purity loss can be associated with the average fidelity of the state to state transformation. 
The fidelity is defined by \cite{glaser2012,rabitz12}:
\begin{eqnarray}
{\cal F}= \frac{1}{2}\left(1+{\cal P}\right)+O(1- {\cal P}). 
\label{avfidpur}
\end{eqnarray}
where $\psi_f$ is the target final state, and $\rho_f$ is the mixed final state attained using noisy controls. 
For high fidelity, i.e., $1-{\cal F} \ll 1$,  one finds:
\begin{eqnarray}
{\cal F}= \frac{1}{2}\left(1+{\cal P}\right)+O\left[(1- {\cal P})^2\right]. 
\label{avfidpur}
\end{eqnarray}

If purity loss is large the average fidelity becomes small and thus
the state-to-state objective is lost.

The instantaneous rate of purity loss for a pure state $\Op \rho=\left|\psi\right\rangle\left\langle \psi\right|$, 
evolving according to Eq.(\ref{liouville}) can be related to the variance of the control operators \cite{viola07}:
\begin{equation}
\dot{\cal P} \equiv -\frac{d}{dt} \bf{Tr}\left\{ \Op \rho^2\right\}|_{\Op \rho=\left|\psi\right\rangle\left\langle \psi\right|}=4  \sum_{k, u_k \neq 0} \Gamma_k(t)\Delta_{\Op X_k}\left[\psi\right], \label{rateofloss}
\end{equation}
where $\Delta_{\Op X_k}\left[\psi\right]$ is the variance of the control operator $\Op X_k$ in the state $\psi$:
\begin{equation}
\Delta_{\Op X_k}\left[\psi\right]\equiv \left\langle \psi\left|\Op X_k^2\right|\psi \right\rangle-\left\langle \psi\left|\Op X_k\right|\psi \label{uncertainty} \right\rangle^2.
\end{equation}
As a result, pure states with larger variance with respect to the control operators $\Op X_k$ will lose purity faster. 
The variance of  a generic state of the system  scales as $N^2$ with the size of the system. 
The scaling of a controlled state variance with $N$ is a central theme in determining the influence of noise. 

The logic of the subsequent presentation is as follows:   First a bound on the time duration of an arbitrary transformations is obtained in the limit of vanishing noise. Next, the lower bound of the purity loss rate for the evolving state in the presence of noise is calculated. For weak noise the two bounds can be combined to obtain the lower bound on the purity loss during the transformation. This bound depends on the relative strength of noise $\eta$. Next,  a class of state-to-state target transformations is defined. The target transformations can be accomplished in the absence of noise due to the assumed state-to-state  controllability of the noiseless system. For \textit{any realization} of the control fields accomplishing the target transformation, the evolving system resides for a long period of time in  states with large variance $\sim N^2$ with respect to the control operators. The large variance will generate  $\sim \eta N^2$ rate of purity loss in the presence of noise in the control fields. It is found that the lower bound on the time of transformation scales as $\sim N^{-1}$, therefore, $\eta$ must be $O(N^{-1})$ in order that the purity loss be negligible. Since, in practice, the relative magnitude of the noise cannot be made arbitrarily small, it follows that the loss of purity cannot be neglected for large systems. As a result, the target transformation  cannot be accomplished, i.e., the state-to-state-controllability is lost in the presence of noise in the control field. 

\subsection{Bounds on minimum control time for obtaining the objective}

We denote the initial and the final (target) states by $| \psi_i \rangle $ to $ | \psi_f \rangle$, respectively. 
Under the assumption that the noise is small, we shall calculate the bound on time $T$ required to perform the transformation $\psi_i \rightarrow \psi_f$ in the zero order in the noise strength. For estimation of this bound an auxiliary operator $\Op A$ is defined such that: 
(i) it commutes with $\Op H_0$; (ii) its expectation value changes during the transformation.  Since $\Op A$ commutes 
with $\Op H_0$  the change of its expectation value during the transformation is due to the operation of the control fields. Accordingly, changes to states that can be reached by free propagation generated by  the (noiseless) drift Hamiltonian $\Op H_0$ will not contribute to the bound on the time.

 We define 
\begin{eqnarray}
&&\Op A=\sum_n s_n \left|n\right\rangle\left\langle n\right|, \\
&&s_n=\bf{sign} \left\{\Delta r_n\right\} \nonumber
\label{Aoperator}
\end{eqnarray}
where $\left|n\right\rangle$    is the drift Hamiltonian $\Op H_0$ basis set and $r_n$ is defined as follows.: In the  $\Op H_0$ basis set the initial state is $\left|\psi_{i}\right\rangle=\sum_n r_{i,n} e^{i \phi_{i,n}}\left|n\right\rangle$
and the final state  is $\left|\psi_f\right\rangle=\sum_n r_{f,n} e^{i \phi_{f,n}}\left|n\right\rangle$. We define a quasi-distance between the two states $|\Delta \bf{r}|$, where:
\begin{eqnarray}
&&\Delta \bf{r}\equiv \bf{r}_{f}-\bf{r}_i, \\
&&\bf{r}_{i}=(r_{i,1},r_{i,2},...) \nonumber \\
&&\bf{r}_{f}=(r_{f,1},r_{f,2},...) \nonumber
\label{deltar}
\end{eqnarray}
Qualitatively, $|\Delta \bf{r}|$ can be viewed as a distance between $\psi_i$  and $\psi_f$ modulo free evolution of the system, i.e., where the distance between any two states, which are connected by the free propagation, is zero.

The change of the expectation value of the operator $\Op A$ during the transformation $\psi_{i} \rightarrow \psi_{f}$ is given by
\begin{eqnarray}
\left\langle\Op A\right\rangle_f-\left\langle\Op A\right\rangle_i&=&\sum_n \left|\Delta r_n \right|\left(r_{i,n} +r_{f,n}\right)\nonumber \\
&\ge& \sum_n \Delta r_n^2=\left\|\Delta \bf{r}\right\|^2, \label{sum2}
\end{eqnarray}
where we have used the fact that the vector of amplitudes $\bf{r}$ is nonnegative, and, therefore, $\left|\Delta r_n\right|>r_{i,n}$ only if $\Delta r_n\ge 0$. 

Inequality (\ref{sum2}) gives the minimal change of the expectation value of the operator $\Op A$ during the transformation $\psi_{i}\rightarrow \psi_f$.
On the other hand, the change of the expectation value of $\Op A$ can be estimated from the Heisenberg equations:
\begin{eqnarray}
\frac{d}{dt}\Op A&=&i\sum_k u_k(t) \left[\Op X_k, \Op A\right]
\end{eqnarray}
where we have used the fact that $\left[\Op H_0,\Op A\right]=0$. Let the time of the transformation be $T$. Then,
\begin{eqnarray}
\left\langle\Op A\right\rangle_f&-&\left\langle\Op A\right\rangle_i=\int_0^{T}\frac{d}{dt}\left\langle \Op A \right\rangle \ dt \nonumber \\
&\le& \sum_k \int_0^{T}\left|u_k(t)\right| \ dt \max_{0\le t\le T}\left|\left\langle \left[\Op X_k, \Op A\right]\right\rangle \right|\  \nonumber \\
&\le& 2  \sum_k \int_0^{T}\left|u_k(t)\right| \ dt \left|\Lambda_{k}\right|, 
\label{minchange3}
\end{eqnarray}
where $\Lambda_k \sim N$ stands for the  eigenvalue of $\Op X_k$, maximal by the absolute value.  In the derivation we have used the fact that the eigenvalues of $\Op A$ are $\pm 1$.   Defining the average control amplitude $\bar{u}_k\equiv \frac{1}{T}\int_0^{T}\left|u_k(t)\right| \ dt$,
 and using Eqs.(\ref{sum2}) and (\ref{minchange3}), we arrive at the inequality
\begin{eqnarray}
T\ge& \left\|\Delta \bf{r}\right\|^2 \left(2\sum_k \bar{u}_k \left|\Lambda_{k}\right|\right)^{-1} \sim \left\|\Delta \bf{r}\right\|^2 \left(2N\sum_k \bar{u}_k\right)^{-1} , \label{timebound2}
\end{eqnarray}
which bounds the time of the transformation for given $\bar{u}_k$. This bound has the structure of a time energy uncertainty relation and is similar to bounds obtained for the transformation to an orthogonal state in Refs.\cite{Margolus98,levitin05,delcampo}. The difference is that the present bound is based on a metric.
If one attempts to reach the objective in a shorter time, fidelity is sacrificed \cite{rabitz12,calarco}.

\subsection{Bounds on purity loss}

The bounds on purity loss are obtained under the assumption that the purity loss $\Delta \cal{P}$ during the transformation is small. 
In this case the evolving state can be approximated by a pure state $\rho(t)=\rho^{(0)}+\rho^{(1)}\approx \rho^{(0)}=\left|\psi(t)\right\rangle\left\langle \psi(t)\right|$.  
Taking the leading contribution of $\rho^{(1)}$ into account, we estimate the lower bound on the purity loss from Eq.(\ref{rateofloss}):
\begin{eqnarray}
\Delta {\cal{P}}&\ge& 4 T\sum_k  \overline{\Gamma}_k  \min_{0\le t \le T}\{\Delta_{\Op X_k}\left[\psi(t)\right]\nonumber \\
&+&\frac{1}{2}T\left\langle \psi(t)\left|[\Op X_k,[\Op X_k,\rho^{(1)}(t)]]\right|\psi(t)\right\rangle\}. \label{minpur}
\end{eqnarray}
where $\overline{\Gamma}_k\equiv T^{-1}\int^T_0\Gamma_k(t)dt$ is the average dephasing rate over the transformation.
We further assume that during the transformation the system follows generic states so that $\Delta_{\Op X_k}\left[\psi(t)\right] \sim (\Lambda_k)^2 \sim N^2$.
In this case, we can neglect the $\rho^{(1)}$-dependent term in the inequality (\ref{minpur}).

 Using the inequality (\ref{timebound2}), we obtain
\begin{equation}
\Delta {\cal{P}}\ge \frac{2 \left\|\Delta \bf{r}\right\|^2   \sum_k   \overline{\Gamma}_k \min_{0\le t \le T}\left\{\Delta_{\Op X_k}\left[\psi(t)\right]\right\}}{\sum_k \bar{u}_k \left|\Lambda_{k}\right|}. \label{minpur2}
\end{equation}

This is a general inequality, applicable to arbitrary initial and final states $\psi_i$ and $\psi_f$. Next we consider a particular class of transformations, where the application of the inequality leads to an explicit scaling relation between the noise strength with the size of the system.

Let us assume that $\psi_i$ and $\psi_f$ are separated by the quasi-distance $|\Delta \bf{r}| \ll 1$. In addition, we assume that (i) $\psi_f$ has  uncertainty $\sim N^2 $ with respect to the control operators  and (ii) any state connected to $\psi_f$ by free evolution has  uncertainty $\sim N^2 $ with respect to the control operators. For example, a generic eigenstate of the free Hamiltonian $H_0$ satisfies assumptions (i)-(ii).

To estimate $\min_{0\le t \le T}\left\{\Delta_{\Op X_k}\left[\psi(t)\right]\right\}$ in (\ref{minpur2}) for the transformation $\psi_i \rightarrow \psi_f$  we find the lower bound on the  variance of $\Op X_k$ in the states $\left|\psi\right\rangle=\sum_n r_n e^{i {\phi}_n}\left|n\right\rangle$  such that $\left\|\bf{r}-\bf{r}_f\right\|\le \left\|\Delta \bf{r}\right\|$.
The variance $\Delta_{\Op X_k}\left[\psi\right]$ is a function of the amplitudes $\bf{r}=(r_1,r_2,...)$ and the phases $\phi_1,\phi_2,...$. 
The free evolution can change the phases at no cost in purity. Therefore, the minimal variance attainable for given amplitudes is sought:
\begin{equation}
\tilde{\Delta}_{\Op X_k}\left(\bf{r} \right)\equiv \min_{\phi_1,\phi_2,...}\left\{\Delta_{\Op X_k}\left[\psi\right]\right\} \label{min}
\end{equation}
By construction, the minimal variance $\tilde{\Delta}_{\Op X_k}\left(\bf{r} \right)$ is a function of $\bf{r}$ only. We shall assume that it is smooth, i.e., for  $|\delta \bf{r}|\le \left\|\Delta \bf{r}\right\|\ll 1$  
\begin{eqnarray}
\tilde{\Delta}_{\Op X_k}\left(\bf{r}_f+\delta \bf{r} \right)=\tilde{\Delta}_{\Op X_k}\left(\bf{r}_f\right)+O(|\delta \bf{r}|) \label{smooth}
\end{eqnarray} 
In view of Eqs. (\ref{min}) and (\ref{smooth}),  inequality (\ref{minpur2}) becomes
\begin{eqnarray}
\Delta {\cal{P}}&\ge&   2 |\Delta \bf{r}|^2  \frac{\sum_k   \overline{\Gamma}_k}{\sum_k \bar{u}_k} \left(\frac{ \min_l\left\{\tilde{\Delta}_{\Op X_l}\left(\bf{r}_f\right)\right\}}{ \max_l\left\{\left|\Lambda_{l}\right|\right\}}\right) \label{gen0}
\end{eqnarray} 
to the leading order in $|\Delta \bf{r}|$.
By assumptions (i)-(ii) above and the definition (\ref{min})  $\tilde{\Delta}_{\Op X_k}\left(\bf{r}_f\right)\sim N^2$. Therefore,
\begin{eqnarray}
 \frac{ \min_l\left\{\tilde{\Delta}_{\Op X_l}\left(\bf{r}_f\right)  \right\}}{ \max_l\left\{\left|\Lambda_{l}\right|\right\}}  =c N,\label{generic}
\end{eqnarray} 
where the number $c$ is of the order of unity. We conjecture that approximation (\ref{generic}) holds for a generic target state $\psi_f$, i.e., assumptions (i)-(ii) above are superfluous.  The reason is that generically  the uncertainty with respect to control operators of a state evolving under the free evolution will remain $\sim N^2$.

Taking (\ref{generic}) into account we put
inequality (\ref{gen0}) into the form

\begin{eqnarray}
\frac{\sum_k   \overline{\Gamma}_k}{\sum_k \bar{u}_k} \le  \frac{\Delta {\cal{P}}}{2 c |\Delta \bf{r}|^2 N} \label{highfid}.
\end{eqnarray}

This inequality holds for $\Delta {\cal{P}}  \ll 1$  and $|\Delta \bf{r}| \ll 1$; the number $c$ is of the order of unity. 
This result, obtained in a less general form in Ref. \cite{khasinkosloff11}, relates the relative noise strength on the controls with the size of the system for a high-fidelity transformation. 

The main result can be stated as follows: For systems and controls defined by the Hamiltonian (\ref{eq:1}) and for a generic state-to-state transformation such that the expectation value of the operator $\Op A$  Eq. (\ref{Aoperator}) changes by $|\Delta \bf{r}| \ll 1$, the purity loss associated with the noise on the controls will be small, $\Delta {\cal{P}} \ll 1$, only if the noise complies with condition (\ref{highfid}). This condition determines the upper bound on the noise strength. For a generic transformation, where the  uncertainty of the evolving state with respect to control operators is $\sim N^2$, the number $c$ is of the order of unity. For fixed change $|\Delta \bf{r}|$ and purity loss $\Delta {\cal{P}}$ the upper bound on the noise strength for a generic transformation will decrease as $N^{-1}$.
For large $N$ the relative noise must decrease indefinitely with the size of the system in order to provide high fidelity.  

\subsection{The noise model}

Typical noise includes a static part and a dynamical part:
\begin{eqnarray}
\Gamma_k(t)=\Gamma_k+c_k u_k(t)^2 \label{noise model}.
\end{eqnarray}
This model reflects the following properties of noise, associated with the control fields: 
\begin{enumerate}
\item{ For a weak field, the dephasing rate $\Gamma_k$ is independent on the amplitude of field but generally depends on a coupling operator, i.e., on $k$.}
\item{ For large amplitude of the control field the noise $\xi_k(t)$ in Eq.(\ref{eq:1}) becomes proportional to the amplitude, $\xi_k(t) \sim u_k(t)$, i.e., the dephasing rate grows as the second power of the amplitude.}
\end{enumerate}

From Eq.(\ref{noise model}) we find 
\begin{eqnarray}
\Gamma_k(t) \ge 2|u_k(t)| \sqrt{\Gamma_k c_k} \label{noise model2}.
\end{eqnarray}
Inserting inequality (\ref{noise model2}) into (\ref{highfid}) we arrive at the necessary condition for controllability:
\begin{eqnarray}
\min_k \sqrt{\Gamma_k c_k} \le \frac{\Delta {\cal{P}}}{2 c |\Delta \bf{r}|^2 N} \label{noise model3},
\end{eqnarray}
for $c\sim 1$, $\Delta {\cal{P}} \ll 1$ and $|\Delta \bf{r}| \ll 1$.

\section{Numerical demonstration}
To test our theory, we consider a control task in the single mode Bose-Hubbard model \cite{gati,gati07,mahan}. The drift Hamiltonian is given by:
\begin{equation}
\begin{array}{lcl}
{\Op{H}_0} =  - \Delta \left( {{\Op{a}}_1^\dag  {\Op{a}}_2  + {\Op{a}}_2^\dag  {\Op{a}}_1 } \right)  +  \frac{U}{2}\left[ {\left( {{\Op{a}}_1^\dag  {\Op{a}}_1 } \right)^2  + \left( {{\Op{a}}_2^\dag  {\Op{a}}_2 } \right)^2 } \right] 
\end{array}
\label{BH}
\end{equation}
where ${\Op a}_i$ the annihilation operator for a particle in the $i$-th well, $\Delta$ is the hopping rate, and $U$ is the strength of the on-site interactions between particles.
The control Hamiltonian is taken as:
\begin{equation}
\Op{X} =  \left( {{\Op{a}}_1^\dag  {\Op{a}}_1  - {\Op{a}}_2^\dag  {\Op{a}}_2 } \right)
\end{equation}

The operators are transformed \cite{mahan} to the $su(2)$ Lie-Algebraic form:
\begin{equation}
{\Op{H}} =  - 2\Delta {\Op{J}}_x  + U{\Op{J}}_z^2 + 2 u(t) {\Op{J}}_z  .
\label{SU2}
\end{equation}
The ${\Op J}_i$ are the operators for the projections of the angular momentum of the $i$ axis, and the Hilbert space of the system of $N$ bosons in this model corresponds to the $J=N/2$ irreducible representation of the $su(2)$ algebra. The addition of the nonlinear term $U{\Op{J}}_z^2$ generates the complete controllability condition since the commutators with the linear terms generate the full $su(N)$ algebra.

The large system limit $N\rightarrow \infty$ is realized by increasing the number of particles in the well, while the group-theoretic structure of the control Hamiltonian (\ref{SU2}) is kept fixed. For a sensible large system limit it is desirable to specify the control task in a group-theoretic  way as well. To this end we shall use a spin-coherent state as an initial state and \textit{a fixed control field.}  Spin-coherent states (SCS) are states having minimal total uncertainty
with respect to the  $su(2)$ generators. The total uncertainty is defined as  \cite{gilmorebook}:
\begin{eqnarray}
\Delta(\psi)\equiv~\sum_{j=1}^3  \Delta_{\Op J_k}\left[\psi(t) \right]
\label{giu},
\end{eqnarray}
and ranges between the total spin number $j$ to $j^2$. Spin-coherent states correspond to $\Delta(\psi)=J=N/2$ and according to Eq. (\ref{rateofloss}) have minimal rate of purity loss. We shall find it more illuminating to use the following related quantity as a measure of uncertainty, the generalized purity of a state \cite{Viola03}: 
\begin{eqnarray}
{\cal P}_{su(2)}[\psi]=\frac{1}{ j^2} \sum_{k=1}^3 \left\langle  \Op X_k\right\rangle_{\psi}^2 ~
\label{ngpurity},
\end{eqnarray}
where  where $j^2$ is the eigenvalue  of the Casimir operator $\Op C=\sum_k^3 \Op  J_k^2$. In view of (\ref{uncertainty}) and since the Casimir operator is group invariant the group-invariant relation of the generalized purity to the uncertainty is
$P_{su(2)}[\psi] =j^2-\Delta(\psi)$ with
 $ 0 \le {\cal P}_{su(2)} \le 1$. Spin coherent states have maximal generalized purity of unity.

We have found numerically for state-to-state transformations 
that control fields that maintain low variance during the evolution (SCS),
are robust to an increase of the size of the Hilbert space. As a result, a seed control field can be easily calculated for a small Hilbert space representation (small $J=N/2$)
and be employed to control with high fidelity for the state-to-state objective for a large Hilbert space (large $J=N/2$) \cite{k262,k276}. 
On the other hand, control tasks that lead to a high variance state or transiently go through a high variance state are extremely difficult to find
i.e. they are hard from the viewpoint of computational complexity.
For example finding the control field leading from a SCS to a cat state is a hard problem.
As a measure of "hardness" we can count the number of iterations required in a monotonic convergent algorithm to reach a pre-specified fidelity.
We found that this number scales at least exponentially with the size of Hilbert space. Moreover even when one additional state is added to
the Hilbert space, the search for the control field has to start from scratch \cite{k262}.

For the numerical demonstration we choose $\Delta = 15$ and $U = 2\Delta/j$. For these parameters the dynamics of the analogue classical system is chaotic. 
The quantum manifestation is that an initially localized state will diverge rapidly and spread over the whole phase space \cite{k276}.

The noise was introduced explicitly as Gaussian white noise in Eq.(\ref{eq:1}) with $\Gamma(t)$ as in Eq. (\ref{noise model}). 
The density operator evolution was traced by solving the Schr\"odinger equation  with many noise realizations and then averaging 
$\Op \rho = \frac{1}{L}\sum_l^L | \psi(\xi_l)\rangle\langle \psi(\xi_l) |$, where $\xi_l$ is the $l$'th realization of the noise.
Up to a hundred realizations were taken for each noise intensity, to limit the determination of the various quantities to less than 5$\%$.
The initial state in all calculations was a SCS which is an eigenstate of the control operator $J_z$, i.e.,  $m_j = j$. 
In this state all the particles occupy a single potential well. 

\begin{figure}[tbp]  
\vspace{-2cm}
\center{\includegraphics[scale=0.4]{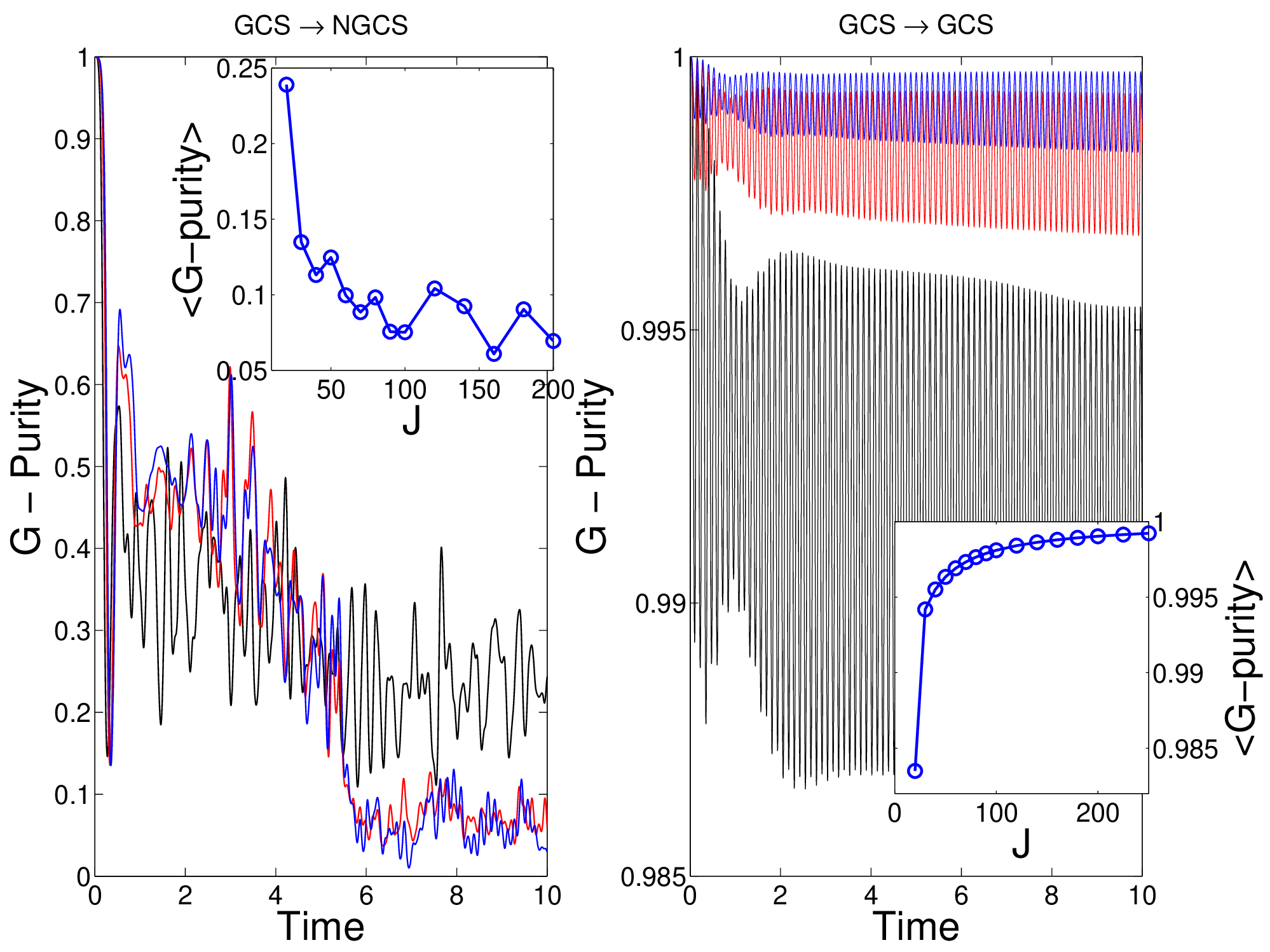}}
 \vspace{-6cm}
 \caption{(color online) Instantaneous normalized generalized purity ${\cal P}_{su(2)}$. 
 (left) Typical temporal values for the generalized purity for a SCS $\to$ non SCS transformation, for $J = 20, 100$, and $200$, 
 in black, red, and blue lines, respectively. (right) Same as the left panel for SCS $\to$ SCS transformations. Insets: 
 time averaged generalized purity ${\cal P}_{su(2)}$ vs. $j$ 
 for the  the SCS $\to$ non-SCS transformations, and for the SCS $\to$ SCS transformations.}
 \label{fig1}
\end{figure}

Two types of state-to-state transformations were examined: (1) a SCS to SCS,  a transformation where the minimum uncertainty with respect to the controls
is maintained during the dynamics, and (2) a SCS to a non-SCS transformation where a large uncertainty develops during the dynamics. 
The same control field with a final target time of $T=10$ was employed for an increasing  Hilbert space dimension.
Type (1) transformation were generated from the temporal local control strategy described in  \cite{k262} which maintain low uncertainty during the evolution.
Figure \ref{fig1} shows that these state-to-state transformations maintain a low uncertainty displayed as high generalized purity, 
${\cal P}_{su(2)}$, Eq. (\ref{ngpurity}), during the whole duration of the evolution. 
This means that the transient state is very close to a SCS.

For  type (2) transformation we chose a random control field of the form:
\begin{equation}
u\left( t \right) = \exp \left[ { - \left( {\frac{{5t}}{T}} \right)^2 } \right]\sum\limits_{k = 1}^{200} {a_k \sin \left( {k\pi t/\tau} \right)} 
\end{equation}
where the random coefficients $a_k$ are picked from a white noise zero to unity distribution. 
The state generated from this random field is designated to be the target
of  state-to-state control. The same control field is used for all calculations with increasing Hilbert space dimension.
The right panel of Fig. \ref{fig1} shows that indeed the ${\cal P}_{su(2)}$ of the transformations generated 
from this field decrease in time and reach a low level typical of a generic target state. 
As expected, the average ${\cal P}_{su(2)}$ of these transformations decreases with increasing Hilbert space size (Cf. insert Fig \ref{fig1}). 
Type (1) and type (2) state-to-state transformations will be employed to study the influence of noise on easy and hard control tasks for a fixed control time.

\begin{figure}[tbp]  
\center{\includegraphics[scale=0.3]{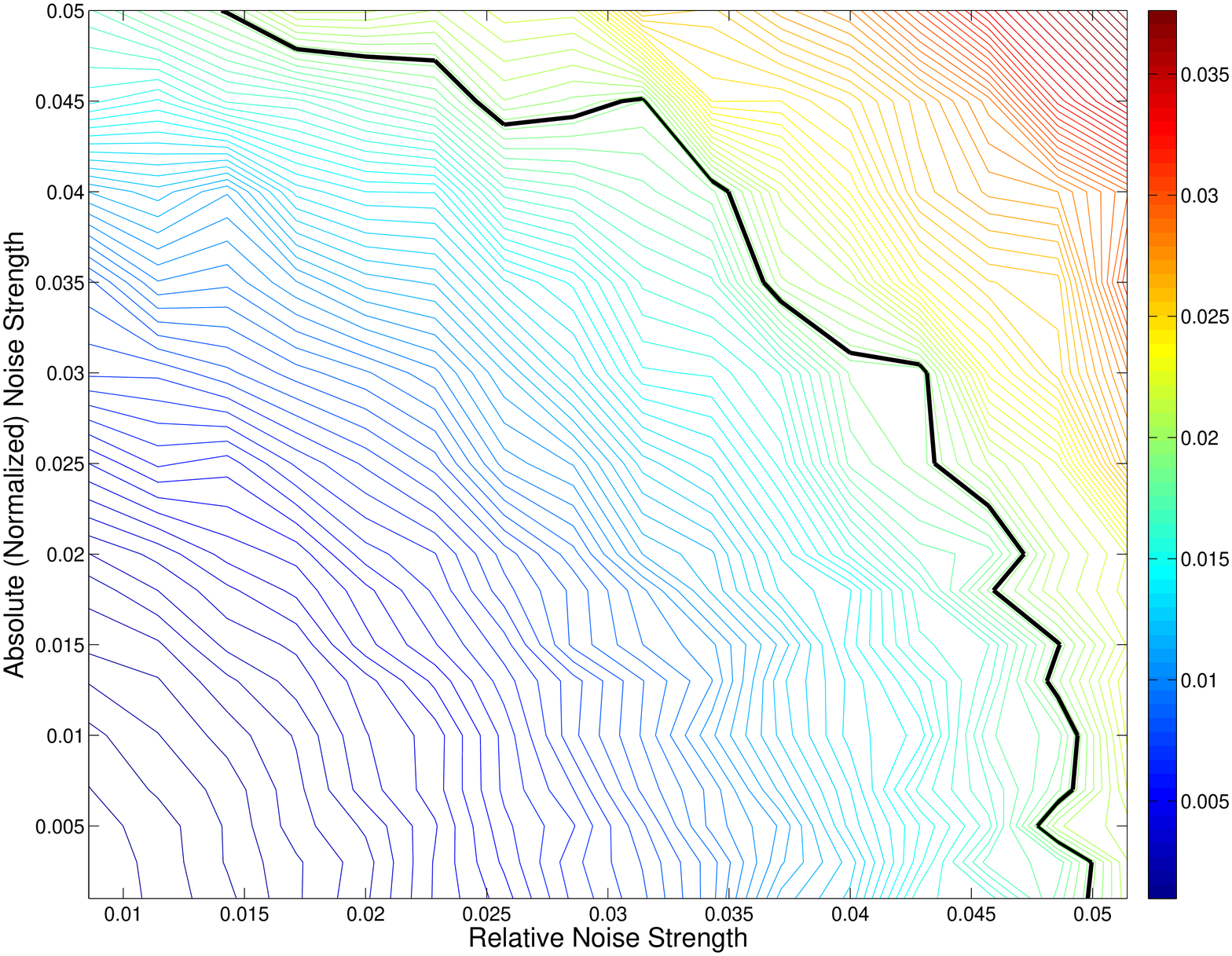}}
 \caption{(color online) Relative and absolute noise role in the dynamics. Deviation from unity of fitness at final time vs. the noise amplitudes, $\Gamma_z$ and $c_z$. The values are given for the type (2) transformation, with $J=20$. The black line on the contour denote a 0.01 degradation of the fitness.}
 \label{figurelN}
\end{figure}

Figure \ref{figurelN} displays a contour plot of the deviation of the fitness from unity at the final time as a function of the relative noise strength $c_z$ and the time-averaged absolute noise $\Gamma_z$, Cf. Eq. (\ref{noise model}). 
The contours in the figure form circles of constant fitness, at least to the accuracy of the stochastic determination of the fitness. 
Seemingly, the two kinds of noise resource have a similar role over the dynamics, and there is no need to consider them separately. 
From this point on, we will set $\Gamma_z = 0$ so that only the relative part of the noise will be taken into account.

\begin{figure}[tbp]  
\center{\includegraphics[scale=0.4]{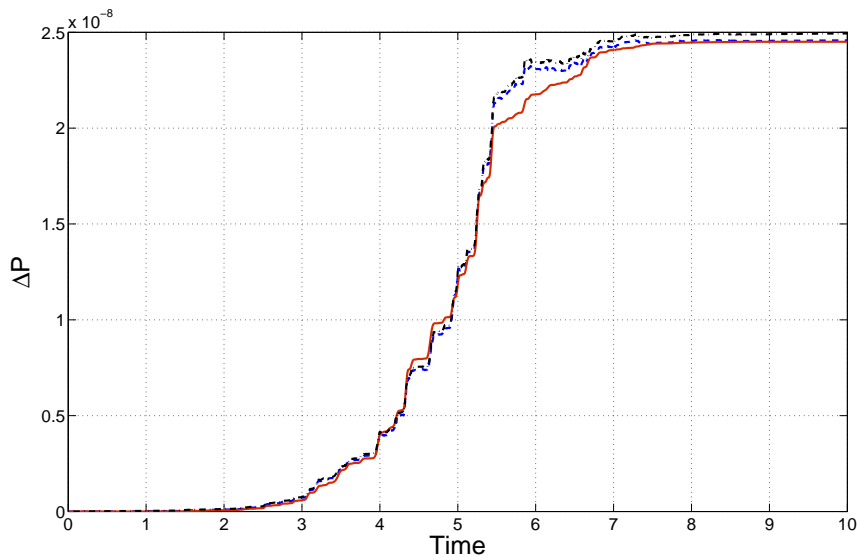}}
\caption{(color online) Typical change in purity $\Delta {\cal P}$ as a function of time (dashed black line). 
The fidelity ${\cal F}$ multiplied by two (blue dotted-dashed line), 
and $\int_0^t \Gamma (t)\Delta_{\hat{J_{z}}}(t) dt $ the trapezoid integration over the control field time the uncertainty (red solid line).
100 realizations of the noise were employed, to reach less than 5$\%$ of error.
The control task is for SCS $\to$ non-SCS transformation, and $J = 20$.}
\label{fig2}
\end{figure}

Figure \ref{fig2} displays a verification of the fundamental relations of this paper, the relation between  fidelity ${\cal F}$
and purity loss $\Delta {\cal P}$, Eq. (4) and the rate of purity loss to the product of noise amplitude and the uncertainty Eq. (5). 
The dashed black line corresponds to the time dependent purity change $\Delta {\cal P}(t)$,  
the blue line corresponds to twice the deviation of the density matrix from (the temporal) perfect fitness ${\cal F}(t)$, 
which is calculated as the scalar product with the temporal state  obtained by the noise free field. 
The red line is a trapezoid integration of the RHS term of eq. (5): $\Gamma \Delta_{\hat{J_{z}}}$. 
Note that the first two quantities $\Delta P$ and ${\cal F}$ were computed by realizations averaging 
the noise to generate the density operator, while the third quantity $\int_0^t \Gamma (t)\Delta_{\hat{J_{z}}}(t) dt $ is calculated for a noiseless control field. 
The comparison is very good despite the noisy averaging.

\begin{figure}[htbp]
\vspace{.1cm}
\center{\includegraphics[scale=0.3]{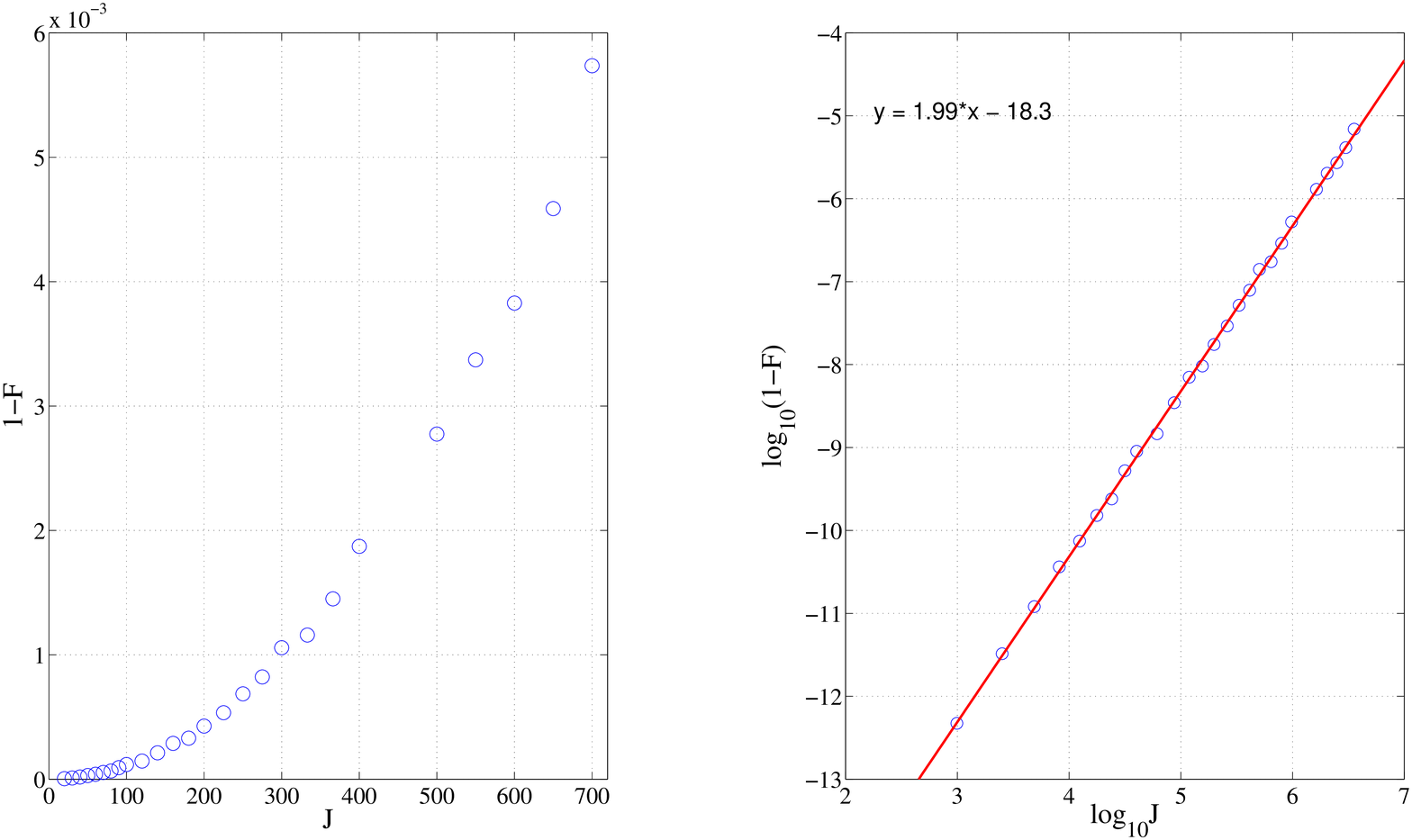}}
\caption{(color online) Final values for the fidelity ${\cal F}$ vs. the Hilbert space size, $J$ in linear and logarithmic scales, for a SCS to Non-SCS task
and set time of 10.}
 \label{fig3}
\end{figure}

\begin{figure}[htbp]  
\center{\includegraphics[scale=0.3]{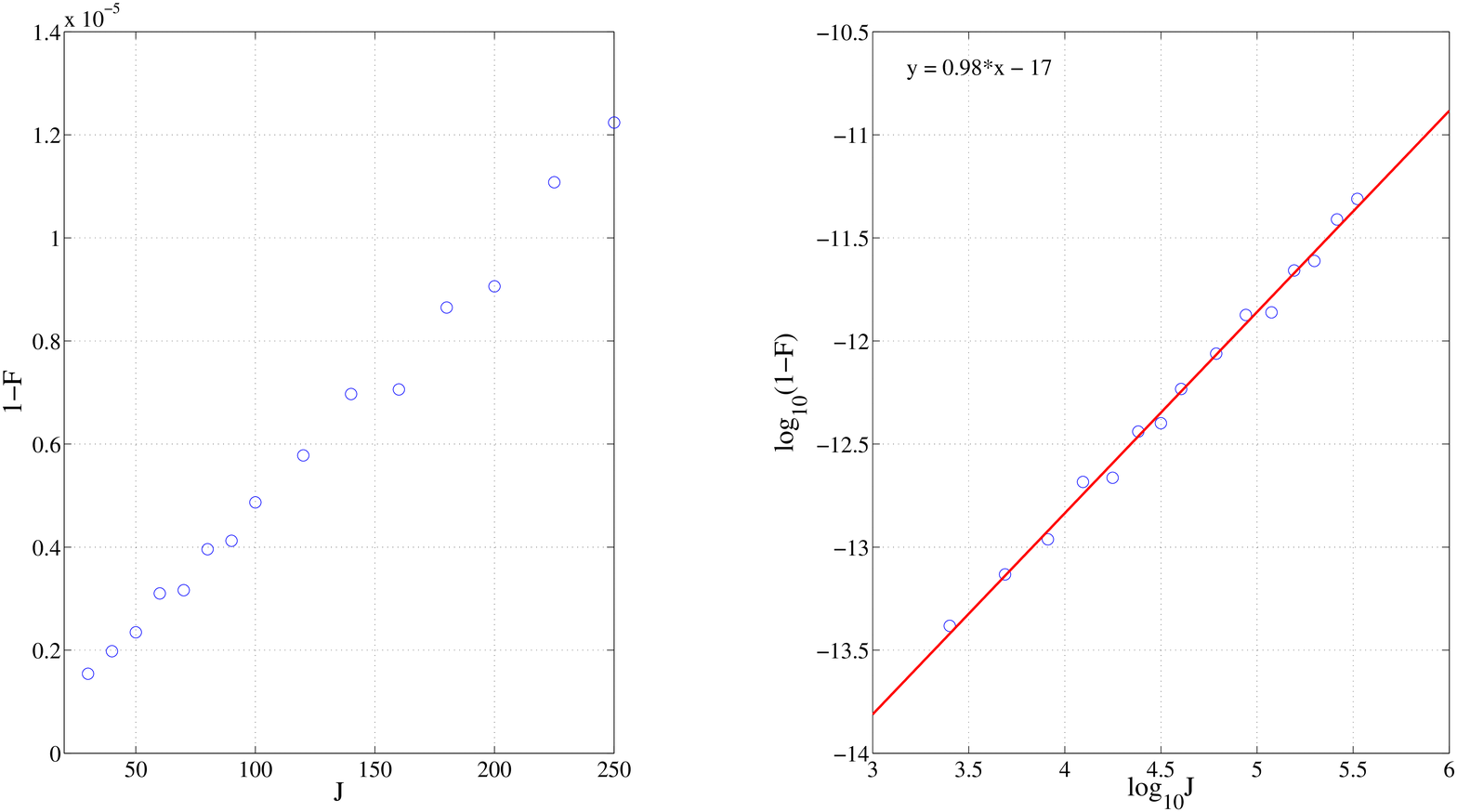}}
\caption{(color online) Final values for  fitness ${\cal F}$ vs.  Hilbert space size, $J$ in linear and logarithmic scales, for a SCS to SCS task
and set time of 10.}
 \label{fig4}
\end{figure}

The scaling of the final fidelity ${\cal F}$ on the size of the Hilbert space was computed for the two types of transformations for a fixed control time. 
Figures \ref{fig3} and \ref{fig4} display the fidelity ${\cal F}$ as a function of $J=N/2$ in linear and in logarithmic scale, 
for  type (1) and (2)  control problems. Examining the logarithmic scale it is obvious that the fidelity ${\cal F}$ for SCS to non-SCS transformation degrades as $N^2$ while the SCS to SCS decrease in  fitness is linear in N. As was discussed at the end of section II D., for the generic state to state transformation the uncertainty of the transient state scales as $N^2$, and so is the fitness and the purity of the transformation. 
In the SCS to SCS transformations the scaling of the uncertainty is linear.

The theoretical minimal  time is inversely proportional to $N$ Cf. Eq. (\ref{timebound2}). The bound is based on the fact
that when the size of  Hilbert space increases so does the maximum eigenvalue of the control operator 
$\Op X_k$ Eq. (\ref{minchange3}), which sets the timescale for control. The reduction of the influence of noise in Eq. (\ref{generic}) compared to Eq. (\ref{rateofloss}), 
is the result of a shorter control time or a smaller control amplitude.
Verification of this scaling relation   is very difficult due to the scaling of computational complexity.
This would require finding a minimum time, optimal control field, for a typical SCS $to$ non-SCS task for each $N$.
From a practical point of view, it is known \cite{k181,k262,calarco,rabitz12} that the computation effort of finding 
a control field increases dramatically for large Hilbert spaces and decreasing designated time.
As was mentioned above, this is a hard task requiring huge computational resources.
As a result, it would be almost impossible to verify computationally the scaling relation of Eq. (\ref{highfid}).

\section{Discussion and Conclusions}

It is impossible to completely eliminate the noise originating from the controller.  This Markovian noise will degrade
the fidelity of achieving the task of state-to-state control. Feedback quantum control \cite{mabuchi}, based 
on a weak quantum measurement of the control operator algebra, will lead to the same result with respect to the measuring operators.
The reason is the equivalence of the noise equation, Eq. (\ref{liouville}), with the Master equation induced by
weak quantum measurement \cite{lajos}.

The sensitivity to noise increases with the computational complexity of the control problem.  In the quantum computation control paradigm the number of gates  increases  with the size of the system. In many other fields, for example in  NMR  \cite{nielsen10} or in control of molecular systems \cite{rice92} a different control paradigm is standard. There the control operators are fixed while the size of the system may vary.  This paradigm was analyzed in the present work. The model studied assumes a finite control algebra with increasing Hilbert space size. The main result is the enhanced sensitivity to control noise when a transient state has a large
variance with respect to the control algebra. A scaling relation between the rate of fidelity loss and the variance is derived and verified numerically. The maximal variance scales as $N^2$ with the size of the system. An upper bound on the strength of tolerable noise, i.e., noise leading to a negligible fidelity loss, is derived for a generic control noise model. The upper bound decreases linearly with the size of the system $N$. In practice, for state-to-state control tasks the noise will dominate, practically scaling as $O(N^2)$ 
 with the size of the system. At best,  according to the upper bound the scaling is $O(N)$. 
 We can speculate that for a control task of generating a unitary transformation, which is equivalent to $N$ simultaneous
 state-to-state control tasks, the decrease in fidelity with noise will practically scale as $O(N^3)$.

The numerical analysis of a model problem associated with the $su(2)$ spectrum-generating algebra verifies the scaling relations and in addition suggests a conjecture that the task of finding a control field that requires a large variance,
for example obtaining a cat state,  is hard. In view of the foregoing discussion this conjecture essentially connects the computational complexity of the control problem with the sensitivity of the controlled system to noise. We believe that this relation is broader than the framework of the control paradigm considered in the present work and applies to any coherent control problem. It seems reasonable to predict that for other control algebras the generalized coherent states (GCS)\cite{perelomov,violabrown} will assume the role of spin coherent states (SGS) considered in the present analysis.
For example,  in the two-dimensional Henon-Heils model with local controls  $\Op X, \Op Y$ and $\Op P_x , \Op P_y$
the GCS are products of single mode coherent states \cite{k276}. Control of a GCS to GCS transformation is easy with respect
to generating a non-GCS state such as an entangled state \cite{viola}. 
It still is to be verified that the noise sensitivity of generating  entangled states is large.  

\subsection*{Aknowledgements}
We want to thank Lorenza Viola, Tomaso Calarco and  Constatine Brief for crucial discussions.
This work is supported by the Israel Science Foundation.
\bibliographystyle{unsrt}

\end{document}